\documentclass[amsmath,twocolumn,superscriptaddress,prl,aps]{revtex4}
\usepackage{amsthm,amsfonts,graphicx,verbatim}
\newcommand{\be}{\begin{equation}}
\newcommand{\ee}{\end{equation}}
\newcommand{\bea}{\begin{eqnarray}}
\newcommand{\eea}{\end{eqnarray}}

\newcommand{\la}{\langle}
\newcommand{\ra}{\rangle}

\renewcommand{\phi}{\varphi}
\renewcommand{\epsilon}{\varepsilon}

\begin{document}

\title{Charge 2{\textit{e}} skyrmions in bilayer graphene}
\author{D. A. Abanin}
\affiliation{Princeton Center for Theoretical Science, Princeton University, Princeton, New Jersey 08544}
\affiliation{Department of Physics, Joseph Henry Laboratories, Princeton University, Princeton, New Jersey 08544}
\affiliation{Kavli Institute for Theoretical Physics, University of California, Santa Barbara, CA 93106}
\author{S. A. Parameswaran}
\affiliation{Department of Physics, Joseph Henry Laboratories, Princeton University, Princeton, New Jersey 08544}
\author{S. L. Sondhi}
\affiliation{Department of Physics, Joseph Henry Laboratories, Princeton University, Princeton, New Jersey 08544}

\date{\today}

\begin{abstract}
Quantum Hall states that result from interaction induced lifting of the eight-fold degeneracy of the zeroth Landau level in bilayer graphene are considered. We show that at even filling factors electric charge is injected into the system in the form of charge $2e$ skyrmions. This is a rare example of binding of charges in a system with purely repulsive interactions.
We calculate the skyrmion energy and size as a function of the Zeeman interaction, and discuss signatures of the charge $2e$ skyrmions in the scanning tunneling microscopy experiments.

\end{abstract}

\maketitle

{\it Introduction.} The four-fold valley and spin degeneracy of Landau levels (LL) in monolayer and bilayer graphene, the recently discovered two-dimensional semimetals~\cite{Novoselov05,Zhang05,Novoselov06}, gives rise to interesting phenomena at high magnetic fields, where the Coulomb interactions between the electrons become important. In the monolayer, the Coulomb interactions lift the LL degeneracy, giving rise to new spin and/or valley polarized incompressible quantum Hall (QH) states~\cite{Zhang06,Abanin07,Jiang07}. The Hamiltonian of the interaction induced quantum Hall states is approximately $SU(4)$ symmetric~\cite{Nomura06,Goerbig06} with respect to the rotations in the combined spin/valley space. The splitting of the LLs thus corresponds to the spontaneous symmetry breaking of the $SU(4)$-symmetric quantum Hall ferromagnet (QHFM). The precise order in which spin and valley degeneracy get lifted is determined by the interplay between the Zeeman interaction and valley anisotropy~\cite{Alicea06,Abanin07-2}, both of which are much smaller than the Coulomb interaction.
The spin- and valley-polarized QH states were predicted~\cite{Yang06} to feature spin and valley skyrmions, which are smooth topologically nontrivial textures of the ferromagnetic order parameter that carry the electron charge $e$~\cite{Sondhi93}. The QHFM states in the monolayer and bilayer graphene are also expected to have interesting edge states properties~\cite{Abanin06,Brey06,Gusynin08}, as well as unusual spectrum of the low-lying collective excitations~\cite{Yang06,Barlas08}. Also, textured states in the vicinity of integer filling factors have been predicted~\cite{Yang06}.

Bilayer graphene features a LL at zero energy, which has a twofold orbital degeneracy: in each valley there are two zero-energy states ($a=0,1$), with wave functions corresponding to the ground state and the first excited state of the magnetic oscillator~\cite{McCann06}. Taking into account valley and spin degeneracies, the zeroth LL in the bilayer is eight-fold degenerate. Coulomb interactions are expected to lift the eight-fold degeneracy~\cite{Barlas08}. In this paper, we consider the interaction induced QH states at even filling factors, and analyze their new properties arising due to the orbital isospin. We shall see that these QH states exhibit interesting collective and topological excitations. We predict that pairs of the excitations of charge $e$ bind into skyrmions that carry charge $2e$. Such binding of charges is surprising, because the Coulomb interactions between electrons are purely repulsive. Another example of such binding was predicted to occur in the spin QHFM with small Zeeman interaction~\cite{Sondhi97,Nazarov98}. The weak pairing of skyrmions considered in Refs.~\cite{Sondhi97,Nazarov98}, however, can occur only when the Zeeman interaction is extremely small; in contrast, charge $2e$ skyrmions in bilayer graphene can be thought of as robust tightly bound pairs, which exist in a wide range of the effective Zeeman interaction.

\begin{figure}
\includegraphics[width=3.2in]{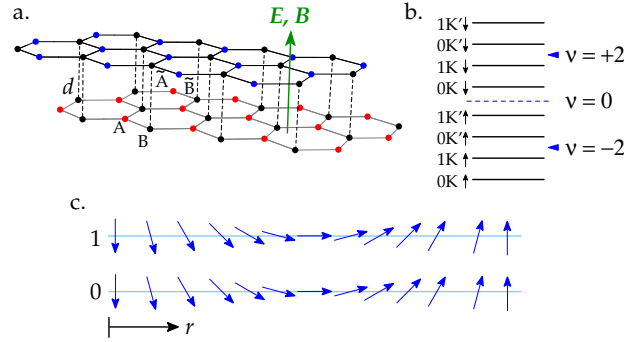}
\label{fig1}
\vspace{-3mm}
\caption[]{(a) Bilayer graphene lattice. 
Perpendicular electric field $E$ generates effective valley Zeeman interaction, $\Delta_v=eEd$, where $d=0.34\,{\rm nm}$ is the separation between the layers. (b) The order of the zeroth LL splitting, assuming that effective valley Zeeman interaction $\Delta_v$ favors $K$ valley states, and $\Delta_v<E_z$. (c) Texture corresponding to the charge $2e$ skyrmion at $\nu=\pm 2$. Vectors illustrate the rotation of the order parameter in the valley space.}
\label{fig0}
\end{figure}

Below we analyze the dependence of skyrmion energy and size on the potential difference between the two layers of bilayer graphene, $\Delta_v=eEd$ (see Fig.~1a), which favors one of the valleys and therefore acts as a valley Zeeman interaction. Owing to the fact that the surface of bilayer graphene is exposed, the skyrmion size can be measured using scanning tunneling microscopy (STM). Furthermore, we find that slightly away from even filling factors, $|\Delta \nu|=|\nu-2M|\ll 1$, there is a finite density of charge-$2e$ skyrmions in the system. At small density, the skyrmions form a triangular lattice, while above a critical density, $\Delta\nu_*$, they form a bipartite square lattice~\cite{Brey95,Abanin09}. The phase transition between the two lattice structures can be observed by STM.



The effective Coulomb interaction Hamiltonian for the zeroth LL in the bilayer is approximately $SU(4)$ symmetric in the valley-spin space, however, the symmetry in the orbital isospin space is broken due to the different orbital wave functions of the two states~\cite{Barlas08}. This results in the following picture of the zeroth LL splitting: at even filling factors ($\nu=2M$ filled sub-levels) $M$ pairs of orbital states with the same valley and spin are filled, while the states at odd filling factors, $\nu=2M+1$ are obtained from the $\nu=2M$ QH state by filling one of the remaining states with orbital isospin $a=0$. This order of the zeroth LL splitting is due to two facts: (i) exchange energy within the LL with isospin $a=0$ is higher than that for the LL with isospin $a=1$; (ii) there is exchange energy between filled $a=0$ and $a=1$ LLs with the same spin and valley, which makes the energy of the state where $a=0,1$ LLs with the same spin and valley are filled (e.g. $0K\uparrow$ and $1K\uparrow$) lower than the energy of a state polarized in the orbital space along $a=0$ direction (e.g., $0K\uparrow$, $0K'\uparrow$ LLs are filled). The order in which valley and spin degeneracies get lifted is determined by the competition between the symmetry-breaking terms: the Zeeman interaction, $E_z=g\mu_B B$, and the effective valley Zeeman interaction $\Delta_v$. In the experiment $\Delta_v$ is typically small, and it can be tuned by gates~\cite{Heersche08}. We assume that $\Delta_v$ is tuned to be smaller than $E_z$. Furthermore, for simplicity, we assume that $\Delta_v$ is small but non-zero and favors the $K$ valley~\footnote{When $\Delta_v$ is extremely small, the ordering in the valley space is determined by the charging energy, which favors states where the charge is distributed equally between the layers. This case will be considered elsewhere~\cite{Abanin09}.}. This leads to the splitting picture illustrated in Fig.~1b. In the following, we shall be especially interested in the states at filling factors $\nu=-2,+2$, marked by arrows in Fig.~1b. Since these two states are related by the particle-hole symmetry, we shall focus on the state at $\nu=-2$.


{\it Landau levels in bilayer graphene.}
We start with recalling the Landau level spectrum in the bilayer graphene~\cite{McCann06}. The low-energy excitations near the $K,K'$ point are described by the Schroedinger equation $\epsilon\psi_{K,K'}=H_{K,K'}\psi_{K,K'}$, with the Hamiltonian given by
\be\label{eq:low_energy}
H_{K,K'}=-\frac{1}{2m} \left[\begin{array}{cc}
         0 & {\pi^\dagger}^2  \\
         {\pi}^2 &  0
      \end{array}
\right] \psi_{K,K'},\quad \pi=p_x- i p_y,
\ee
where the effective mass $m$ can be expressed in terms of the interlayer hopping amplitude $\gamma_1\approx 0.39\,{\rm eV}$ and the Fermi velocity in the monolayer $v_F\approx 10^6\, {\rm m/s}$, $m=\gamma_1/2v_F^2$. For the $K$  valley the upper (lower) component of the wave function corresponds to the amplitudes on the sublattices $A(\tilde B)$ (see Fig.~1a), which belong to different layers.
For the $K'$ valley the order of components is reversed, such that the upper (lower) component corresponds to the $\tilde{B}(A)$ amplitude.

To analyze the LL spectrum, we choose the Landau gauge, $A_y=Bx$, $A_x=0$, for which the eigenstates can be classified according to the value of the wave vector $k_y$, $\psi_{K,K'}(x,y)=e^{ik_y y}\psi_{K,K'}(x)$. The wave vector $k_y$ translates into the guiding center position, $X=k_y\ell_B^2$, where $\ell_B=\sqrt{\hbar c/eB}$ is the magnetic length. Below for simplicity we shall choose units where $\ell_B=1$. The effective 1d Hamiltonian for $\psi_{K,K'}(x)$ takes the following form,
\be\label{eq:hamiltonian_magnetic}
H_{K,K'}=-{\hbar\omega_c} \left[\begin{array}{cc}
         0 &  a_X^2 \\
         {a^\dagger_X}^2 &  0
      \end{array}
\right], \quad a_X=i(\partial_{x}+(x-X)),
\ee
where $\omega_c=eB/mc$ is the cyclotron energy. The Hamiltonian (\ref{eq:hamiltonian_magnetic}) has two zero modes with the following wave functions,
$\psi_{K,K'}^{a}(x,y)=e^{iXy} (0, \phi_{a,X}(x)), \quad a=0,1, $
where $\phi_{a,X}(x)$ denotes the $a$-th excited level of the magnetic oscillator. Below we shall denote the annihilation operators of the zero modes by $c_{a,\kappa,X}$, $\kappa=K,K'$.

{\it Coulomb interaction.}
Now we proceed to the analysis of the zeroth LL splitting.
We neglect the LL mixing (the effects of LL mixing will be considered elsewhere~\cite{Abanin09}), which allows us to project the Coulomb Hamiltonian onto the zeroth LL. The interaction Hamiltonian can be written in the following form,
\be\label{eq:coulomb}
H_{int}=\frac{1}{2S}\sum_{{\bf q}\kappa  \kappa ' } V({\bf q})\rho_{\kappa}({\bf q})\rho_{\kappa '}(-{\bf q}),
\ee
where $\rho({\bf q})$ are the density operators, $S=L_xL_y$ is the sample volume, $\kappa,\kappa'$ are the valley indices, and the matrix element is given by $V({\bf q})=\frac{2\pi e^2}{\epsilon q}$~\footnote{In principle, the intra- and inter-valley components of the Coulomb interaction are slightly different, $V_{\kappa,\kappa}({\bf q})=\frac{2\pi e^2}{\epsilon q}$,  $V_{\kappa,\kappa'}({\bf q})=\frac{2\pi e^2}{\epsilon q} e^{-qd}$ ( $\kappa\neq \kappa'$), which is due to the fact that states in different valleys reside in different layers, separated by the distance $d\approx 0.34\, {\rm nm}$; however, since $d\ll \ell_B$ for typical fields $B=10\,{\rm T}$, the difference between intra- and inter-valley interactions is small and can be neglected.}.

In order to project the Coulomb interaction onto the zeroth LL, we introduce projected density components as follows,
\be\label{eq:density}
\rho_{\kappa}({\bf q})=\sum _{a,b} F_{ab}({\bf q})\bar \rho_{\kappa}^{ab}, \,\, \bar\rho _{\kappa}^{ab}({\bf q})=\sum_{\bar X}\exp\left(iq_x\bar X\right) c_{a,\kappa,X_-}^\dagger c_{b,\kappa,X_+}.
\ee
were $X_{\pm}=\bar X\pm \frac{q_y}{2}$, and $F_{00}({\bf q})=e^{-q^2/4}$, $F_{11}({\bf q})=(1-{q}^2/2)e^{-q^2/4}$ are the usual form-factors for the lowest LL and the first excited LL, and
$F_{01}({\bf q})=-\frac{q_y+iq_x}{\sqrt 2}e^{-q^2/4}$, $F_{10}({\bf q})=\frac{q_y-iq_x}{\sqrt 2} e^{-q^2/4},$
are the form-factors corresponding to the density components which mix the two orbital states. The effective Coulomb interaction within the zeroth LL is obtained by plugging Eq.(\ref{eq:density}) into Eq.(\ref{eq:coulomb}).


{\it Nature of the states with even filling factors.}
We now analyze the order in which the eight-fold degeneracy of the LL gets lifted.
The split QH states with filling factor $|\nu| \leq 3$ ($\nu+4$ filled sub-levels) correspond to the following wave functions,
\be\label{eq:wave_function}
|\Psi_{\nu}\ra=\prod_{i=1}^{\nu+4} \prod_X d_{i,X}^\dagger |\Omega\ra,
\ee
where $d^\dagger$ are linear combinations of the $c^\dagger$ operators:
\be\label{eq:a}
d_{i,X}^{\dagger}=\sum_{a,\kappa,s} {\bar U^i_{a,\kappa,s}} c_{a,\kappa,s,X}^{\dagger},
\ee
where $\bar U$ is a unitary matrix, and $s$ is the electron spin.

To find the ground state for $\nu=-2$, we compare the energies of two states: (i) two $a=0$ LLs with different valley and/or spin indices are filled, for example, $d_1^\dagger=c_{0,K,\uparrow}^\dagger$, $d_2^\dagger=c_{0,K',\uparrow}^\dagger$, and (ii) $a=0$ and $a=1$ LLs with the same valley and spin indices are filled, $d_1^\dagger=c_{0,K,\uparrow}^\dagger$, $d_2^\dagger=c_{1,K,\uparrow}^\dagger$.
The energy of the first state is twice the exchange energy of a non-degenerate lowest LL,
\be\label{eq:E_1}
\la H_{int} \ra_1=-2N\Delta_0, \,\,\,\, \Delta_0=\frac{1}{2}\sqrt{\frac{\pi}{2}} \frac{e^2}{\epsilon\ell_B}
\ee
where $N$ is the total number of states in one non-degenerate LL.
Averaging the effective Coulomb interaction over the second state, we obtain
\be\label{eq:E_2_2}
\la H_{int} \ra_2=-\frac{11}{4}N\Delta_0.
\ee
Thus the energy of the second state is lower than the energy (\ref{eq:E_1}) of the first state, and the spin- and valley-polarized state $|\psi_0\ra$ is the ground state at $\nu=-2$. The state at $\nu=+2$ can be obtained from $|\psi_0\ra$ by charge conjugation.


{\it Charge 2e skyrmions.}
Now we proceed to discussing excitations of the $\nu=-2$ QH state. 
 The lowest-energy electron-hole pair at $\nu=-2$ is obtained by removing an electron with orbital isospin $a=1$ from the filled LL, and putting it into one of the empty LLs. The energy of such excitation, $E_{eh}=\frac{7}{2}\Delta_0$, is lower than the energy $E_{eh}'=4\Delta_0$ of a particle-hole excitation that is obtained by removing an electron with isospin $a=0$.

In some QHFMs, the lowest energy charge excitations are skyrmions, which are topologically nontrivial smooth textures of the order parameter~\cite{Sondhi93}. On the qualitative level, the textures carry charge because the charge and spin (and/or valley) dynamics in the QHFM are entangled~\cite{Sondhi93, Moon94}.

Can skyrmions exist in bilayer graphene? Skyrmions of charge $e$ are energetically unfavorable because they involve flipping valley isospin (or spin) for either $a=0$ or $a=1$ states in some region, and in that region the filled $a=0$ and $a=1$ states would have different valley isospin (spin), which leads to a loss of the exchange energy $\Delta_0$ per flipped valley isospin.

Another possibility is skyrmions of charge $2e$, which can be created by making two identical valley textures for $a=0$ and $a=1$ orbital states. Such textures are described by a unit vector ${\bf n}$, with $n_z=-1(+1)$ corresponding to filling $K(K')$ states. On the intuitive level, we expect such textures to be energetically favorable: since $a=0$ and $a=1$ orbital states rotate simultaneously, no exchange between $0$ and $1$ states is lost. Below we find the energy of the $2e$ skyrmion, and, by comparing it to the energies of the single-particle excitations, establish that such skyrmions are indeed energetically favorable.

Before we proceed to the quantitative analysis of charge $2e$ skyrmions, we would like to compare excitations at the even and odd filling factors. For simplicity, let us consider the excitations of the state $\nu=-3$, which corresponds to filled $0K\uparrow$ LL. The lowest energy electron-hole pair is obtained by removing an electron from the $0K\uparrow$ LL and putting it in the $1K\uparrow$ LL; owing to the exchange between $0K\uparrow$ and $1K\uparrow$ states, the energy of such excitation, $E_{eh}^{odd}=\Delta_0$, is lower than the energy $\tilde{E}_{eh}^{odd}=2\Delta_0$ of an excitation where the excited electron resides in a LL with a different valley/and or spin index.
The existence of orbital skyrmions at $\nu=-3$ is unlikely, because such skyrmions correspond to filling $a=1$ states in some region, which leads to a loss of the exchange energy equal to $\Delta_0/4$ per flipped orbital isospin.

{\it Skyrmion energy.}
In order to compute the energy of charge $2e$ skyrmion, we derive an effective Hamiltonian describing textures of the order parameter,
\be\label{eq:texture}
|\psi\ra=e^{-i{\hat{O}}}|\psi_0\ra.
\ee
In our analysis, we follow the microscopic approach developed in Ref.~\cite{Moon94}; as we shall see below, the dynamics of the order parameter in the bilayer graphene is richer than that in the case of $SU(2)$ and $SU(4)$-symmetric QHFM, owing to the presence of the orbital degree of freedom.
We parametrize the rotation operator $\hat{O}$ as follows,
\be\label{eq:O}
\hat{O}=\sum_{{\bf q},a,b,\mu} \Omega_{ab}^{\mu}({\bf q}) \hat{S}^{\mu}_{ab}(-{\bf q}),
\ee
\be\label{eq:S}
\hat{S}^{\mu}_{ab}(-{\bf q})=\sum_{\bar{X}} e^{iq_x \bar{X}} \frac{\tau^{\mu}_{\kappa \kappa '}}{2} c^\dagger_{a,\kappa,X_-} c_{b,\kappa', X_+},
\ee
where $\tau$ are the Pauli matrices. The rotation (\ref{eq:O}) is described by four complex parameters, \be\label{eq:parameters}
u_a=\Omega_{aa}^x+i\Omega_{aa}^y, a=0,1, \,\, v=\Omega_{10}^x+i\Omega_{10}^y, \,\, w=\Omega_{01}^x+i\Omega_{01}^y.
\ee
The parameters $u_{0} (u_1)$ correspond to rotations that involve $0K$ and $0K'$ ($1K$ and $1K'$) states, while $v, w$ parametrize rotations which transform $0K$ into $1K'$, and $1K$ into $0K'$, and vice versa. To simplify calculations, we assume that the rotations are small ($|u_a|\ll 1, |v|\ll 1, |w|\ll 1$).
Then we can expand the texture energy $E=\la \psi_0| e^{i\hat{O}^\dagger} H e^{-i\hat{O}}|\psi_0 \ra -\la\psi_0| H |\psi_0 \ra$ in series in the powers of $\hat{O}$. Doing that and evaluating the mass and stiffness terms, we obtain~\cite{Abanin09},
\begin{widetext}
\begin{eqnarray} \nonumber
 && E_0= -\frac{\rho_0}{2}\int \frac{dz d\bar z}{2i} [  4\bar{\partial} u_0^* \partial u_0  +7 \bar{\partial} u_1^* \partial u_1+6\{ \bar{\partial} u_0^* \partial u_1+c.c.\}
+2\sqrt{2}\{u_0^*(\partial v+\bar{\partial} w)+c.c.\}+{\sqrt{2}}\{ u_1^*(\partial v+\bar{\partial} w)+c.c.\}  \\ \label{eq:energy2}  &&
-2\{\bar\partial v^* \partial v+\bar\partial w^* \partial w \}-{3}\{\bar\partial v^* \partial w+c.c.\}+2(u_0^*-u_1^*)(u_0-u_1)+4 v^* v+ 3w^* w ], \,\,\, \rho_0=\frac{e^2}{16\sqrt{2\pi} \kappa\ell_B},
\end{eqnarray}
\end{widetext}

We are interested in the low-energy excitations, where $a=0$ and $a=1$ states are rotated simultaneously in the valley space (see Fig.1c). This corresponds to setting $u_0=u_1=u$, which ensures that the mass term that contains $u_0, u_1$ components in Eq.(\ref{eq:energy2}) vanishes. Minimizing the energy with respect to $v,w$ yields
$v=\frac{3\sqrt{2}}{4}\bar\partial u+..., \,\, w=\sqrt{2}\partial u+...,$ where ellipsis denotes  the higher order gradient terms. Notice that $v,w$ are proportional to the gradients of $u$, which implies that $v,w$ are much smaller than $u$ for the case of slowly varying textures.
Substituting the expression for $v,w$ into Eq.(\ref{eq:energy2}), we obtain the gradient term for $u$,
\be\label{eq:energy_u}
E_{st}=-2{\rho_s}\int \frac{dz d\bar z}{2i} \bar\partial u^* \partial u,\,\,\, \rho_s=\frac{25}{8}\rho_0.
\ee
For what follows, it is convenient to rewrite the stiffness energy (\ref{eq:energy_u}) in terms of the $O(3)$ order parameter ${\bf n}=(-u_y, u_x, 0)$,
\be\label{eq:energy_n}
E_{st}=\frac{\rho_s}{2}\int d^2r \, (\partial_{\mu} {\bf n})^2.
\ee
Although we have derived the above equation assuming that ${\bf n}$ deviates slightly from ${\bf n}=(0,0,-1)$, due to the rotational invariance in the valley space Eq.(\ref{eq:energy_n}) is valid for an arbitrary slowly varying configuration of the order parameter.

As our next step, we evaluate the charge density of the texture, $\delta \rho=\la \psi_0| e^{i\hat{O}^\dagger}\hat\rho e^{-i\hat{O}}|\psi_0 \ra-\la\psi_0| \hat\rho |\psi_0 \ra$, where $\hat\rho$ is the density operator. We find that the charge density is twice the Pontryagin index density,
\be\label{eq:top_density}
\delta\rho({\bf r})=2e p({\bf r}), \,\,\, p({\bf r})=-\frac{1}{8\pi}\epsilon_{\mu\nu}({\bf n}[\partial_{\mu}{\bf n}\times \partial_{\nu}{\bf n}]).
\ee
Notice that this relation differs from the usual $SU(2)$ QHFM case~\cite{Sondhi93,Moon94} by a factor of 2, which corresponds to the fact that the texture rotates states in both $a=0,1$ LLs.

Apart from the stiffness term (\ref{eq:energy_n}), there are two other contributions to the texture energy: the valley Zeeman term and the long-range Coulomb interaction,
\bea
\label{eq:valley_zeeman}
&&H_{z}=\Delta_v n_0 \int d^2{\bf r}\, n_z, \\
\label{eq:skyrmion_coulomb}
&&
H_{coul}=\frac{1}{2}\int d^2{\bf r}  d^2{\bf r'} \delta \rho({\bf r}) \frac{1}{|{\bf r}-{\bf r'}|} \delta\rho({\bf r'}),
\eea
where $n_0=1/2\pi\ell_B^2$ is the LL density of states.

The simplest topologically nontrivial texture of the order parameter ${\bf n}$ has topological charge $1$ and an electric charge $\pm 2e$. This is to be contrasted with the usual skyrmions~\cite{Sondhi93}, which carry charge $\pm e$.
In the limit of vanishing $\Delta_v$, the Coulomb repulsion (\ref{eq:skyrmion_coulomb}) forces skyrmions to be infinitely large, $l_s\to\infty$, where $l_s$ is the skyrmion size. Then the skyrmion energy is determined solely by the stiffness term,
\be\label{eq:skyrmion}
E_{sk}=4\pi \rho_s=\frac{25}{16}\Delta_0.
\ee
The energy of the skyrmion-antiskyrmion pair, $2E_{sk}=25\Delta_0/8$, is lower than the energy of two electron-hole pairs, which equals $7\Delta_0$. Therefore, in the limit $\Delta_v\to 0$, pairs of electron (hole) excitations bind into charge $2e$ skyrmions (antiskyrmions).

At finite $\Delta_v$ the skyrmion size is determined by the competition between the effective valley Zeeman and Coulomb energies~\cite{Sondhi93}. Optimizing the skyrmion energy with respect to its size, we find with logarithmic precision,
\be\label{eq:skyrmion_size}
\frac{l_s}{\ell_B}\approx \left( \frac{9\pi^2}{32} \right)^{1/3} \tilde\Delta_v ^{-1/3} |\log\tilde\Delta_v |^{-1/3},
\ee
where $\tilde{\Delta}_v=\Delta_v/(e^2/\ell_B)$. The skyrmion energy is increased compared to the case $\Delta_v=0$,
 $
 E_{sk}(\tilde\Delta_v)=\frac{25}{16}{\Delta_0}+A \Delta_0 \tilde\Delta_v^{1/3}|\log \tilde\Delta_v|^{1/3}$, where $A=\frac{3^{4/3}\pi^{5/6}}{2^{11/6}}.$

{\it Experiment.}
We now briefly address experimental manifestations of the charge $2e$ skyrmions. The most direct way to observe the skyrmions in by the STM, which allows one to study the properties of the individual skyrmions, as well as the skyrmion configuration at finite density. For an individual skyrmion, STM may be used to study the dependence of the skyrmion size (\ref{eq:skyrmion_size}) on the valley Zeeman interaction; the latter can be tuned by gates. Furthermore, in the vicinity of even filling factors, $|\nu-2M|\ll 1$, there is a finite density of skyrmions in the system. When the skyrmion density $\tilde n$ is low, such that the distance between skyrmions is much larger than the skyrmion size (\ref{eq:skyrmion_size}), $\tilde n^{-1/2}\gg a$, the skyrmions form a triangular Wigner crystal. At larger densities, $\tilde n^{-1/2}\sim a$, when the skyrmions start to overlap, we find~\cite{Abanin09} that, similarly to the spin QHFM case~\cite{Brey95}, the skyrmions rearrange into a square lattice. Such behavior will result in a periodic modulation of the local density of states, which can be measured by an STM and used to determine the skyrmion lattice symmetry.

{\it Acknowledgements.} We thank Philip Kim, Jens Martin, and Amir Yacoby for helpful discussions. This research was  supported in part by the National Science Foundation under Grant No. PHY05-51164 (DA).










\end{document}